\documentstyle[12pt,pc96]{article}
\begin{document}

\title{Metastable Fractal Aggregates as a Result of
Competition Between Diffusion-Limited Aggregation and Dissociation}

\author{Yuriy G. Gordienko and Elena E. Zasimchuk}

\institute{Institute of Metal Physics of the National Academy of Sciences\\
	36 Vernadsky St., Kiev 252142, Ukraine\\
           {\em email:\/} {\tt gord@imp.kiev.ua}}

\maketitle

\begin{abstract}
The cellular automaton model is used to simulate diffusion
and aggregation with dissociation of point particles in 2D.
A continuous phase transition is found that separates creation of compact
aggregates and fractal ones. The transition is the
function  of  pair-interaction energy ($E_b$),
type of neighborhood and temperature $T$.
Manifestations of the transition  in  real  physical systems
are discussed.
\end{abstract}

\section{Introduction}

In the modern theory of fractal growth
the significant efforts were dedicated to
defining models (\cite{DLA}, \cite{DBM}) and
developing the theoretical concepts necessary
to understand these model and to compute their properties
analytically \cite{DLAReview}. The crucial elements of
the simulation models and theoretical schemes are:

- irreversibility;

- screening and freezing;

- spontaneous evolution to a fractal structure.\\
Moreover, the fractal properties are presumed to be well defined only
asimptotically with respect to the large time and scale limits.
It means that the problem of fractal growth consists in
considering regions of
the system that are very far from growing interface and that will
not continue to be modified by further growth.
But there are many experimental evidents of surface
aggregates of different morphologies: fractal and dendritic
\cite{Brune}, compact and fractal \cite{Roder}, convex
and concave ones \cite{Con}, which were
determined by {\em kinetics} of 2D growth processes.
Recent simulations brought to light
dependence of aggregate morphology from coverage \cite{Amar}
in irreversible diffusion-limited aggregation (DLA).
These observations of transition from fractal
to compact aggregates can be explained by deposition of atoms
in internal regions of the aggregates.
But the passage from {\em irreversible} to {\em reversible}
aggregation was also achieved by tuning a value of pair bond energy
\cite{Smilauer}.  It points out an influence of dissociation processes on
morphology of aggregates.

Investigation of morphology as a function of deposition
conditions during DLA with dissociation (DLAD) is of great interest
in the spirit of the great importance of building
low-dimensional nanostructures with desirable geometry.
That is why in the following the cellular automaton (CA) model
of metastable fractal aggregates created by DLAD
is considered (Sec.2). In Sec.3 the dependence of fractal
dimensions as a function of kinetics of DLAD is presented
and the results are discussed in Sec.4.

\section{Cellular automaton model of diffusion-limited aggregation
with dissociation}

The rules of our CA model are similar to the classic DLA model \cite{DLA}
with the following addition: particles are randomly generated beyond
a growing aggregate with a constant rate $F$ and diffuse with a
diffusion rate $D \sim \exp(-E_m/k_{\rm{B}}T)$, where $E_m$ is
a migration energy.
And they can dissociate from an aggregate with the following rate
$R \sim D_e \exp(-\sum_{NNs}E_b/k_{\rm{B}}T)$, where $E_b$ is
a pair bond energy, $D_e \sim \exp(-E_e/k_{\rm{B}}T)$ is a diffusion
rate of particles in aggregate
(for example, in the simplest case $D=D_e$ \cite{Smilauer}).
Local interaction with the nearest neighbors is taken into account
by summation. Our simulations involved a square lattice with
sizes ranging from $(100 \times 100)$ to $(600 \times 600)$ with
periodic boundary conditions. Simulations were conducted for
different values of the ratio $\kappa = \exp((E_b+E_m-E_e)/k_{\rm{B}}T)$
and the ratio $D/F$ ranging from $10^4$ to $10^9$.
The CA evolves in discrete time  steps  and  the
state  of  each  site  at  the  next  time  step  is
determined from the state of the site itself  and  those
of  the nearest-neighbor sites (NNs). The von Neumann neighborhood
(with four NNs) and Moore one (with eight NNs and diagonal coupling)
were used.

\section{Metastable fractal aggregates with quasi-stable fractal dimensions}

\begin{figure}[bht]
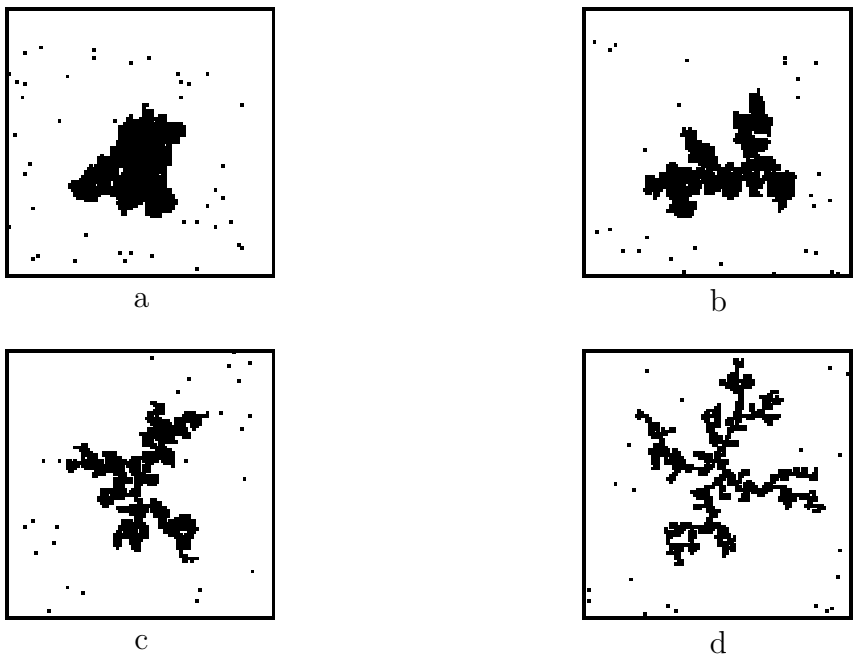

\centering
\begin{minipage}{5.9cm}
\input{FIG1A.PIC}
\end{minipage}
\hspace{1.5cm}
\begin{minipage}{5.9cm}
\input{FIG1B.PIC}
\end{minipage}

\vspace{1cm}

\begin{minipage}{5.9cm}
\input{FIG1C.PIC}
\end{minipage}
\hspace{1.5cm}
\begin{minipage}{5.9cm}
\input{FIG1D.PIC}
\end{minipage}

\vspace{0.5cm}

\caption{Typical morphologies created by diffusion-limited
aggregation with dissociation (DLAD) on initial stages
for the von Neumann neighborhood
at $D/F = 10^6$ for ($100 \times 100$) sections:
a) $\kappa = 20$; b) $\kappa = 100$; c) $\kappa = 200$; d) $\kappa = 1000$.}
\label{Patterns}
\end{figure}

Fig.\ref{Patterns} illustrates typical morphologies
for the von Neumann neighborhood on initial stages at $D/F = 10^6$ for
($100 \times 100$) sections: a) $\kappa = 20$; b) $\kappa = 100$;
c) $\kappa = 200$; d) $\kappa = 1000$. Fractal dimensions were determined from
fits to the linear region in log-log plots, which were obtained by
density-density correlation function ($d_c$) and box counting measurement
($d_b$). For this purpose a central part of an aggregate was considered as
an inscribed square of maximum dimension.
On initial stages of aggregation (the number of particle in aggregate
$n < 10^2$)
length scale is very small for reliable calculation of fractal dimensions.
In the range
$10^2 < n < 10^3$ the dimensions grow and achieve asymptotic values for
$n \approx 10^3$.
The values of $d_c$
shown in Table \ref{vonNeumData} for the von Neumann neighborhood
 and Table \ref{MooreData} for the Moore neighborhood
with the main statistical characteristics of interpolation lines.
Box counting measurements
demonstrate several linear regions in log-log plots: from initial
region with $d_b = 2$ to a staircase part with alternating regions of
$d_b \approx 1$
and $d_b \approx 2$. The regions have small length scales for reliable
determination of $d_b$. Although a core of aggregates becomes compact with
time, they continue to generate "DLA-like" branches.
When deposition was ceased, aggregates underwent significant restructions.
The previous simulations
shown that aggregate splitted in many small parts and coalesced
with creation of
one large compact aggregate with $d_c = 2$.

\section{Simulated phase diagrams as a new tool
for investigation of surfaces}

\begin{table}
\caption{Fractal dimensions ($d_c$)
for the von Neumann neighborhood
obtained by density-density correlation measurements.
Mean values of the
interpolation lines are denoted by word "mean", standard deviations
are shown in row "sd" and correlation coefficients
are given in row "corr".}
\label{vonNeumData}

\begin{center}
\begin{tabular}{|c|c|c|c|c|c|c|c|c|c|}
\hline
$\kappa$   	& 20   	& 40   	& 100  	& 150  	& 200   	& 400   	& 1000  & 2000\\
\hline
mean    & 1.956 & 1.869 & 1.857 	& 1.794 & 1.755 & 1.669  & 1.653 & 1.652\\
\hline
sd      	& 0.012 & 0.014 & 0.005 & 0.008 & 0.003 & 0.007  & 0.004 & 0.006\\
\hline
corr	& 0.9998 & 0.9997 & 0.9999 & 0.9999 & 0.9999 & 0.9996 & 0.9999 & 0.9996\\
\hline
\end{tabular}
\end{center}
\end{table}

\begin{table}
\caption{Fractal dimensions  ($d_c$)
for the Moore neighborhood
obtained by density-density correlation measurements.
Mean values of the
interpolation lines are denoted by word "mean", standard deviations
are shown in row "sd" and correlation coefficients
are given in row "corr".}
\label{MooreData}

\begin{center}
\begin{tabular}{|c|c|c|c|c|c|c|c|c|c|}
\hline
$\kappa$  & 4 	& 10 	& 20   	& 40   	& 100  	&  200 	& 400 \\
\hline
mean     & 1.945	& 1.862 & 1.776 & 1.71 & 1.708 	& 1.652 & 1.662\\
\hline
sd      	& 0.007 & 0.007 & 0.003 & 0.007 & 0.006 & 0.018	& 0.027\\
\hline
corr	&0.9997& 0.9998&0.9999&0.9997&0.9997& 0.9992& 0.9980\\
\hline
\end{tabular}
\end{center}
\end{table}

In the well-known work of Witten and Sander \cite{DLA}
DLA with sticking probability,
which depends on the number of bonds to filled
neighbor sites $B$,
was considered {\em without} dissociation.
Particle can stick to an aggregate only during
a fraction $s$ of its encounters: $s = t^{3-B}$, where $t<1$.
In addition to their conclusion that "long-range
properties seem to be unaltered" for $0.1<t<0.5$ we
should like to emphasize the crucial changes in DLA
caused by dissociation.
It should be noted that the whole aggregate can split
in unlinked parts, but scale invariance defined by
density-density correlations remains.

The aforementioned results allow to propose the new
method for investigation of adatom-adatom and adatom-surface
interactions by means of
investigation of scale invariant aggregates.
For this purpose careful simulations of DLAD
for different deposition rate, $\kappa$, type of
interaction are necessary.
It will give model phase diagrams for any kind of
real aggregating adatoms and substrate.
Testing the real aggregation processes for different
deposition conditions and comparing theirs with the
simulated model phase diagrams will allow to bring
to light new features of local interaction at surfaces.
We used the naive Arrhenius-type expessions for dissociation
rates. It is obvious the more realistic scheme will need
the more sophisticated algorythms and codes with more reliable
results.

In addition to this we recently observed the very complicated band
patterns in metals during plastic
deformation of metals \cite{Glas}. They can be explained  by  development  of
disturbances  of concentration of point-type defects and their
agglomerates \cite{GordienkoPhil}.
The crucial point of this assumption is based on
possibility of point defects to agggregate
in scale invariant structures with different dimensions and
consequently different abilities to capture point-type defects.
Fractal dimensions of aggregates are dependent
on pair interaction energy and they define
the selection rate among heterogeneities with further band pattern formation.
The extended version of aforementioned investigation in 3D will also
allow to make clear mechanisms of
pattern formation in plastically deformed metals.

\paragraph{Acknowledgments.\/}
The authors gratefully acknowledge Prof.A.A.Snarski for
helpful comments on this work.

\end{document}